\documentclass[useAMS]{mn2e}
\usepackage{lscape}
\usepackage{rotating}
\usepackage{amssymb}

\def\msun{\hbox{M$_\odot$}}

\def\t4{\hbox{t$_{\rm 4}$}}

\def\reff{\mbox{$R_{\rm eff}$}}		
\def\sbmag{\hbox{mag arcsec$^{-2}$}}	
\newcommand\phn{\phantom{0}}		

\def\kms{\hbox{km$\,$s$^{-1}$}}
\def\cm3{\hbox{cm$^{-3}$}}

\input psfig.sty
\input epsf.sty
\usepackage{graphicx}
\usepackage{epsfig}
\title[Cluster profiles and sizes in NGC~7252]
{Luminosity profiles and sizes of massive star clusters in NGC~7252}
\author[N. Bastian et al.] {N. Bastian$^{1,2}$, F. Schweizer$^{3}$, P. Goudfrooij$^{4}$, S.S. Larsen$^{5}$, \& M. Kissler-Patig$^{6,7}$\\
$^1$ Astrophysics Research Institute, Liverpool John Moores University, Egerton Wharf, Birkenhead, CH41 1LD, UK \\
$^2$ Excellence Cluster Universe, Boltzmannstr. 2, 85748 Garching, Germany\\
$^3$ Carnegie Observatories, 813 Santa Barbara Street, Pasadena, CA 91101, USA \\ 
$^4$ Space Telescope Science Institute, 3700 San Martin Drive, Baltimore, MD 21218, USA\\
$^5$ Department of Astrophysics / IMAPP, Radboud University Nijmegen, P.O. Box 9010, 6500 GL Nijmegen, The Netherlands\\
$^6$ European Southern Observatory (ESO), Karl-Schwarzschild-Strasse 2, 85748, Garching, Germany\\
$^7$ Gemini Observatory, 670 N. A'ohoku Place, Hilo, Hawaii, 96720, USA \\
}
\date{Accepted. Received; in original form}
\pagerange{\pageref{firstpage}--\pageref{lastpage}}
\pubyear{2012}
\begin{document}
\maketitle
\label{firstpage}
\begin{abstract}
We present {\em Hubble Space Telescope (HST)} Wide-Field Camera 3 (WFC3) images of the merger
remnant NGC~7252.  In particular, we focus on the surface brightness profiles and effective radii \reff\ of $36$ young massive clusters (YMCs) within the galaxy.  All the clusters have masses exceeding $10^5$\,\msun\ and are, despite the 64~Mpc distance to the galaxy, (partly) resolved on the {\em HST\,} images.  Effective radii can be measured down to $\sim2.5$~pc, and the largest clusters have \reff\ approaching $20$~pc.  The median \reff\ of our sample clusters is $\sim6-7$~pc, which is larger than typical radii of YMCs ($\sim2.5$~pc). This could be due to our sample selection (only selecting resolved sources) or to an intrinsic mass--radius relation within the cluster population.  We find at least three clusters that have power-law profiles of the Elson, Fall, \& Freeman (1987, ``EFF'') type extending out to $\ga\,$150~pc.  Among them are the two most massive clusters, W3 and W30, which have profiles that extend to at least 500 and 250~pc, respectively.  Despite their extended profiles, the effective radii of the three clusters are 17.2, 12.6 and 9.1~pc for W3, W26 and W30, respectively.  We compare these extended profiles with those of YMCs in the LMC (R136 in 30~Dor), the Antennae galaxies (Knot S) and in the nearby spiral galaxy NGC~6946.  Extended profiles seem to be a somewhat common feature, even though many nearby YMCs show distinct truncations.  A continuous distribution between these two extremes, i.e. truncated or extremely extended, is the most likely interpretation.  We suggest that the presence or absence of an extended envelope in {\em very young clusters} may be due to the gas distribution of the proto-cluster giant molecular cloud, in particular if the proto-cluster core becomes distinct from the surrounding gas before star formation begins.

\end{abstract}

\section{Introduction}
\label{sec:intro}

NGC 7252 is a prototypical merger remnant that hosts one of the largest populations of young massive
star clusters known (Whitmore et al.~1993; Miller et al.~1997, hereinafter M97), including two clusters with stellar masses exceeding $10^7$\,\msun\ (Maraston et al.~2004, hereinafter M04; Bastian et al.~2006, hereinafter B06).  While there is ongoing star-formation within a central disk of gas (Schweizer~1982) the cluster population as a whole is dominated by an extended halo population of YMCs with an age of $\sim400\pm100$~Myr (e.g., Schweizer \& Seitzer 1998, hereinafter S98).

Studies of luminosity profiles of YMCs have shown that they often differ from those of classic globular clusters. The latter are generally well fit by King (1962) profiles that display a truncation at large radii.  This truncation is thought to be related to the tidal radius, inside which stars are bound to the cluster, while outside the stars move according to the galactic potential.  Many YMCs, on the other hand, do not display such a truncation, but rather are well fit by extended power-law envelopes.  This was quantified by Elson, Fall, \& Freeman~(1987, hereinafter EFF) for YMCs in the Large Magellanic Cloud.  These authors fit profiles of the form $I( r ) = I_0(1 + r^2/a^2)^{-\gamma/2}$, where $r$ is the distance from the cluster centre and $a$ is a characteristic radius.  For $r \gg a$ the EFF profile becomes a simple power-law.  EFF profiles have been shown to also provide good fits to YMCs in M33 (San Roman et al.~2012) and in spiral galaxies more generally (Larsen~2004).

Observations of young cluster systems have shown that the clusters themselves are often grouped into larger structures, cluster complexes, with radii of tens to hundreds of parsecs (e.g.,~Zhang et al.~2001; Larsen~2004).  These complexes often appear to be centrally concentrated, hosting a dominant cluster in the centre, with a similar EFF-type power-law decrease in surface brightness with increasing radius (Bastian et al.~2005).  The outer envelope of the complexes is, at least partly, made up of smaller individual clusters.  Fellhauer \& Kroupa~(2005) carried out $N$-body model simulations of a system of massive star clusters distributed as a Plummer sphere, assuming that the clusters are in virial equilibrium, and found that the clusters merge within a few hundred Myr.  The remnant formed in their simulations has a massive central cluster and an extremely extended outer halo, leading the authors to suggest that the massive cluster W3 in NGC~7252, may have formed in such a manner.

Interested by the extreme nature of the YMCs in NGC~7252, i.e. their high masses and relatively large ages ($\sim400$~Myr), we have obtained {\em HST}/WFC3 images of NGC~7252 to study the luminosity profiles of these clusters in detail.  Despite the relatively large (64 Mpc for $H_0 = 75$ \kms\ Mpc$^{-1}$) distance to the host galaxy, a large number of the clusters are (semi)resolved.  We were able to trace the profiles of three of the most massive clusters out to distances of more than 150~pc from the cluster centres.  Specifically, our main goal was to see whether the cluster profiles are truncated and, if not, how far they extend.  We were also interested in studying whether the presence or absence of any extended envelope might correlate with the projected distance of clusters from the remnant's centre, which would be expected if the tidal truncation of clusters in the galaxy potential would already have had an observable impact.


The present paper is organized as follows.  In \S~\ref{sec:obs} we present the data and methods used to create the point spread function, and then outline the steps taken to generate the cluster sample.  In \S~\ref{sec:profiles} we introduce the techniques used to measure the effective radii of the clusters and study the luminosity profiles of three massive clusters (W3, W26 and W30) in detail.  In \S~\ref{sec:radii} we study the distribution of \reff\ for the entire population and investigate the possibility of a mass--radius relation for massive clusters.  In \S~\ref{sec:reff_dgc} we investigate the relation between the effective radius and galactocentric distance of the clusters in our sample.  In \S~\ref{sec:discussion} we compare the extended profiles of clusters in NGC~7252 with those of other known clusters and discuss the possible origin of such extended envelopes.  Finally, \S~\ref{sec:conclusions} presents our conclusions.


\section{Observations}
\label{sec:obs}

\subsection{WFC3 observations}

Broad-band images of NGC~7252 were obtained with {\em HST}/WFC3 on 2010 July 29 through the $F475W$ (SDSS $g$) and $F775W$ (SDSS $i$) filters (GO-11554, PI: N.\ Bastian).  The filters were chosen to provide coverage in the blue and red part of the optical, and to be wide in order to maximise the throughput and signal-to-noise ratio of the brightness profiles.  

The total exposure time through each filter was  2120~s, split into four
500 s exposures plus two short, 60 s exposures to avoid saturation of the
bright central regions of each cluster and the galaxy.  The individual
exposures in each filter
were taken with small spatial offsets (``dithers'') between one another to
allow improved spatial sampling of the point spread function (PSF) as well as
the elimination of hot pixels and cosmic-ray hits during image combination.
Prior to running the {\tt PyRAF/stsdas} task {\tt MultiDrizzle} to produce 
a combined image for each filter, we created sky variance maps for each
individual exposure for the purpose of deriving weight maps for the final
image combination.  These maps were constructed from the sky values of the
individual exposures, the flatfield reference file, the dark current reference
file, and the read-out noise values as listed in the image headers.  As such,
these sky variance maps contain all noise components of the image except
for the Poisson noise associated with the sources on the image. 

The final run of {\tt MultiDrizzle} was performed by shrinking the input
pixels by $20$\% at the stage where input pixels are drizzled onto the
output image grid (i.e., a value of 0.8 was adopted for the {\tt final\_pixfrac} parameter), and we chose an output
image pixel size of 0\farcs028/pixel.  These various parameters were selected after
extensive experimentation, and they allow a good match to the degree of
subsampling induced by the dither pattern we used. 
Saturated pixels in the 500 s exposures were replaced by the corresponding pixels
in the short exposures while running {\tt MultiDrizzle}.  This was achieved by
setting the appropriate data quality flag for the affected pixels of the long exposures.  



Additionally, we used WFC3 $F336W$ ($U$) images (GO-11691, PI: P.\ Goudfrooij) that will be presented in more detail in Goudfrooij et al. (in prep.).  In brief, the total $F336W$ exposure was 3050 s, and the images were reduced in much the same way as those described above.  These $U$ images were not used for size or profile determination, but only for photometric selection.

\subsection{Other imaging}

In addition to the WFC3 images of NGC~7252, we also used archival data to study the luminosity profiles of other YMCs.  In particular, we used {\em VLT}/HAWK-I near-infrared $K_{s}$-band images of 30~Doradus to study the outer profile of the central cluster, R136 (programme ID 60.A-9283).  The data were taken as part of the commissioning of the instrument and are presented in more detail in Campbell et al.~(2010).

For the YMC in NGC~6946, we used the {\em HST}/WFPC2 $V$-band images presented in detail by Larsen et al.~(2001).  Finally, for the study of Knot~S in the Antennae galaxies we used {\em HST}/ACS $V$-band images taken from the {\em Hubble Legacy Archive}.  These images are presented in more detail in Bastian et al.~(2009) and Whitmore et al.~(2010).


\subsection{Creation of a model PSF}

In order to get accurate measurements for the profiles and radii of the target clusters we required an accurate model of the point-spread function (PSF) of our data.  We did this in the following two ways.

The first was to select sources of various brightnesses throughout the image that appeared to be unresolved (i.e., they all had the same FWHM) and use them in the {\em IRAF} task {\em PSF} to generate an empirical PSF.  For this we used $14$ sources located throughout the image.  The disadvantage of this method is that it will blur any effect of the PSF varying across the chips.


A second set of PSFs was created from the grid of empirical WFC3 PSFs
assembled by Jay Anderson (hereafter, {\it ePSFs}; see Anderson \& King~2006 for the ACS version). Since
the latter PSFs are created for individual {\tt *\_flt.fits} images rather
than for images combined with {\tt MultiDrizzle}, we wrote a package based on
Maurizio Paolillo's {\it 
Multiking\/}\footnote{http://www.na.infn.it/$\sim$paolillo/Software.html}
suite of scripts (modified to work for WFC3 images). The package creates 
empirical PSFs at the location of each GC candidate in blank versions 
of each individual input {\tt \_flt.fits} WFC3 image, and then combines those
files by {\tt MultiDrizzle} in the exact same way as the final NGC~7252
images. These PSFs were then subsampled by a factor 10 in order for them to be used
appropriately within {\it ISHAPE\/} (see below), using polynomial
interpolation.

We compared the empirically derived PSFs with the PSFs in the {\it ePSF} grid by deriving the radii of the spectroscopically confirmed clusters (see \S~\ref{sec:sample}) using {\em ISHAPE} (Larsen 1999; see also \S~\ref{sec:ishape} below) with both sets of PSFs.  We did not find any systematic differences between the results using either of the PSFs, with measured FWHMs being the same to within $\sim10$\%.  This difference is smaller than that due to the choice of the radius over which to measure the size, or the choice of the type of profile to fit (King vs EFF).  For the results presented below we adopt the {\it ePSF} grid. 

For each of the methods we treated the two filters ($F475W$ and $F775W$) independently.

\subsection{The cluster sample}
\label{sec:sample}

The main goal of this work was to measure the surface-brightness profiles (SBPs) and \reff\ of the clusters within NGC~7252, focussing on the $\sim400\pm100$~Myr old population formed during the galaxy merger, i.e., clusters outside the central 12\arcsec-diameter star-forming disk (M97, S98).  This population is known to include extremely massive clusters (M97, S98, M04, B06), even though the population is consistent with a power-law mass function of index $-2$ (B06).

\begin{figure}
\includegraphics[width=8.5cm]{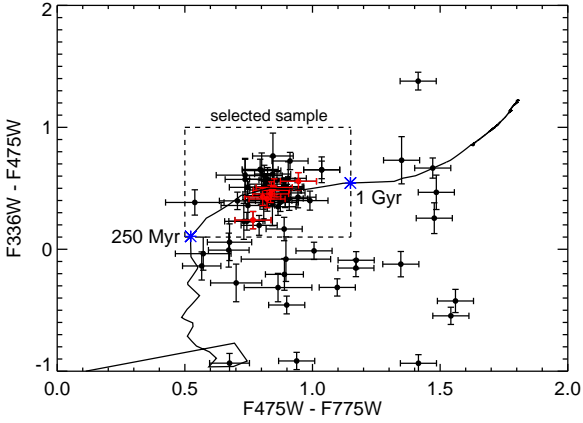}
\caption{The colour--colour diagram of isolated sources in NGC~7252.  Red points represent clusters that have been spectroscopically confirmed, while the dashed box indicates the colour selection applied.  All sources within the box are considered cluster candidates.  The solid line represents the GALEV SSP models for solar metallicity, and the limiting ages of the models that pass our colour criteria are given in the panel.}
\label{fig:colour}
\end{figure}

In order to study the SBPs of the clusters, high S/N is required.  We selected $\sim100$ source candidates ($18 < F475W < 25$~mag) outside the central disk (galactocentric radii $>4.2\arcsec$, $\sim1.3$~kpc), extending out to 108\arcsec\ ($\sim33$~kpc) from the galaxy center.  We excluded candidates with nearby neighbours that would complicate the SBP analysis.  In order to remove background galaxies we measured the magnitudes and colours of each of these candidates from the $F336W$, $F475W$, and $F775W$ images, using an aperture of 10 pixels in radius.  The resulting colour--colour diagram is shown in Fig.~\ref{fig:colour}, where the spectroscopically confirmed clusters (S98) are shown in red.  The dashed box shows the colour region where we selected our sample.  Additionally, we show the evolutionary track of a GALEV simple stellar population (SSP) model of solar metallicity as a solid line (Kotulla et al.~2009).  The limiting ages of the model clusters that pass our colour selection are 250~Myr and 1~Gyr, as marked.  Note, however, that the observed cluster colours are concentrated around 400\,--\,500~Myr.  The applied colour selection removes background galaxies and effectively limits our study to the intermediate age ($\sim400$~Myr) cluster population.  Using our colour selection criteria, we found 52 candidate clusters.

\begin{figure}
\includegraphics[width=8.5cm]{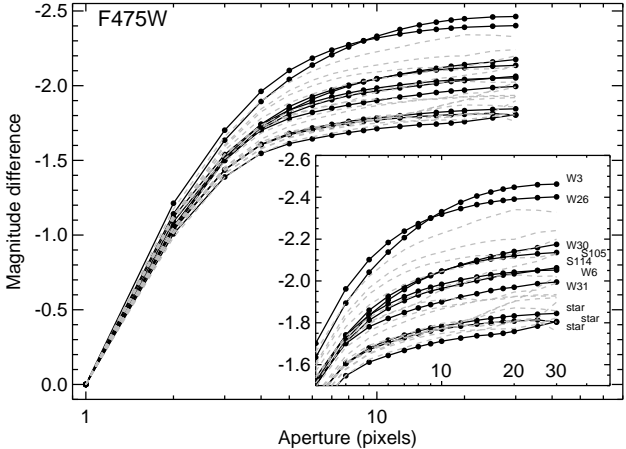}
\caption{Curves of growth, normalised to an aperture of 1 pixel, as a function of aperture radius for a sample of sources in the WFC3 FOV (before any colour selection has been applied).  The inset shows a zoom in on the end of the distribution, with confirmed clusters and likely stars labelled.  Confirmed clusters are shown in black with filled circles, while other sources in the field are shown as dashed grey lines.}
\label{fig:growth}
\end{figure}

We then fit these candidates with {\em ISHAPE}, adopting an EFF profile with $\gamma=3.0$, and measured their concentration index (CI; see \S~\ref{sec:ishape} below).   We also performed aperture photometry for each of the candidates with aperture radii ranging from 1 to 30~pixels.  Examples of the resulting growth curves are shown in Fig.~\ref{fig:growth}.

Using these two methods, we selected objects that appeared extended in each method.  Figure~\ref{fig:comp} shows a comparison of the measured CIs and \reff\ for an EFF profile with $\gamma=3.0$.  The dashed lines mark limits where we can resolve sources.  Red triangles represent confirmed clusters within NGC~7252 based on optical spectroscopy and agreement with the systemic velocity (S98).  Additionally, we performed resolution tests by adding artificial clusters (with known radii) to the images which were analysed in the same way as discussed above, in order to confirm the adopted limits.

As Fig.~\ref{fig:comp} clearly shows, {\em all of the confirmed clusters from S98 are resolved}.
Also, of the total sample of 52 candidate clusters 36 are resolved in both the {\em ISHAPE} and CI analyses.  However, we note again that our cluster sample is not complete in any way, as only bright and resolved clusters were selected.

Table~\ref{tab:catalogue} gives the IDs, magnitudes, CIs, estimated effective radii (adopting an EFF profile with $\gamma=3.0$) and galactocentric distances of all the clusters that pass our colour selection and appear resolved using both {\em ISHAPE} and the CI method.


\section{Surface-brightness profiles and radii of the massive clusters}
\label{sec:profiles}

\subsection{{\em ISHAPE} and the Concentration Index}
\label{sec:ishape}

Once the cluster candidates were selected we used three methods to study their profiles and effective radii.

First, for the brightest clusters (e.g., W3, W26 and W30) we performed a detailed analysis of their SBPs with the intention to constrain the profile shape, effective radius and extent to which their envelopes reach.  These will be discussed individually below.  

As a second method, we used the {\em ISHAPE} algorithm (Larsen~1999) with a fixed profile shape (an EFF profile with $\gamma = 3$) to measure each cluster's FWHM, which in turn can be used to derive its \reff.  The fitting radius was chosen to be 20 pixels (= $0.56\arcsec \approx 175$~pc), although we did not find any systematic effects as a function of the fitting radii.

As a third method, we measured for each cluster the concentration index CI (e.g., M97; Whitmore et al.~2010), which is defined as the magnitude difference between measurements with a 1 pixel and a 3 pixel radius aperture.  We also experimented with different aperture combinations and found consistent results.

In all plots of the SBPs to be shown below, we have added a constant to the derived SBP so that the surface brightness becomes 0~\sbmag\ in the first bin.  When profiles of more resolved clusters are shown for comparison, we shift these to have a value of 0~\sbmag\ at the same physical radius as that for the NGC~7252 clusters.

\begin{figure}
\includegraphics[width=8.5cm]{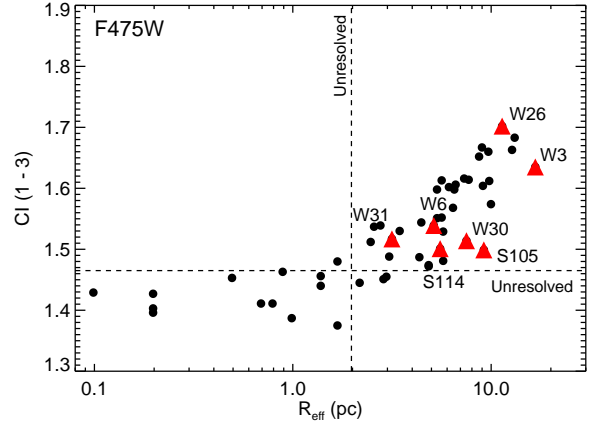}
\caption{Estimated effective radii \reff\ determined by using {\em ISHAPE} (adopting an EFF profile with $\gamma=3$) are plotted versus concentration indices CI(1$-$3) measured with apertures of one and three pixel radius.  The vertical dashed line marks the FWHM of 0.2 pixels used for the adopted profile, while the horizontal dashed line denotes the cut applied between point-like and extended sources.  The spectroscopically confirmed clusters are shown as (red) triangles.  Only sources that passed our colour selection are shown.}
\label{fig:comp}
\end{figure}

\subsection{W3}
\label{subsec:W3}

\subsubsection{Profile and effective radius}

Cluster W3 is an enormously massive cluster of $\sim 8\times 10^7$\,\msun\ (S98, M04).  In order to determine its profile and effective radius we fit various profile types with {\em ISHAPE}, letting the FWHM as well as the index $\gamma$ (for an EFF profile) or concentration parameter $c$ (for a King profile) vary.  We tested multiple initial guesses (index and concentration) for the algorithm and found good convergence.

\begin{figure}
\includegraphics[width=8.5cm]{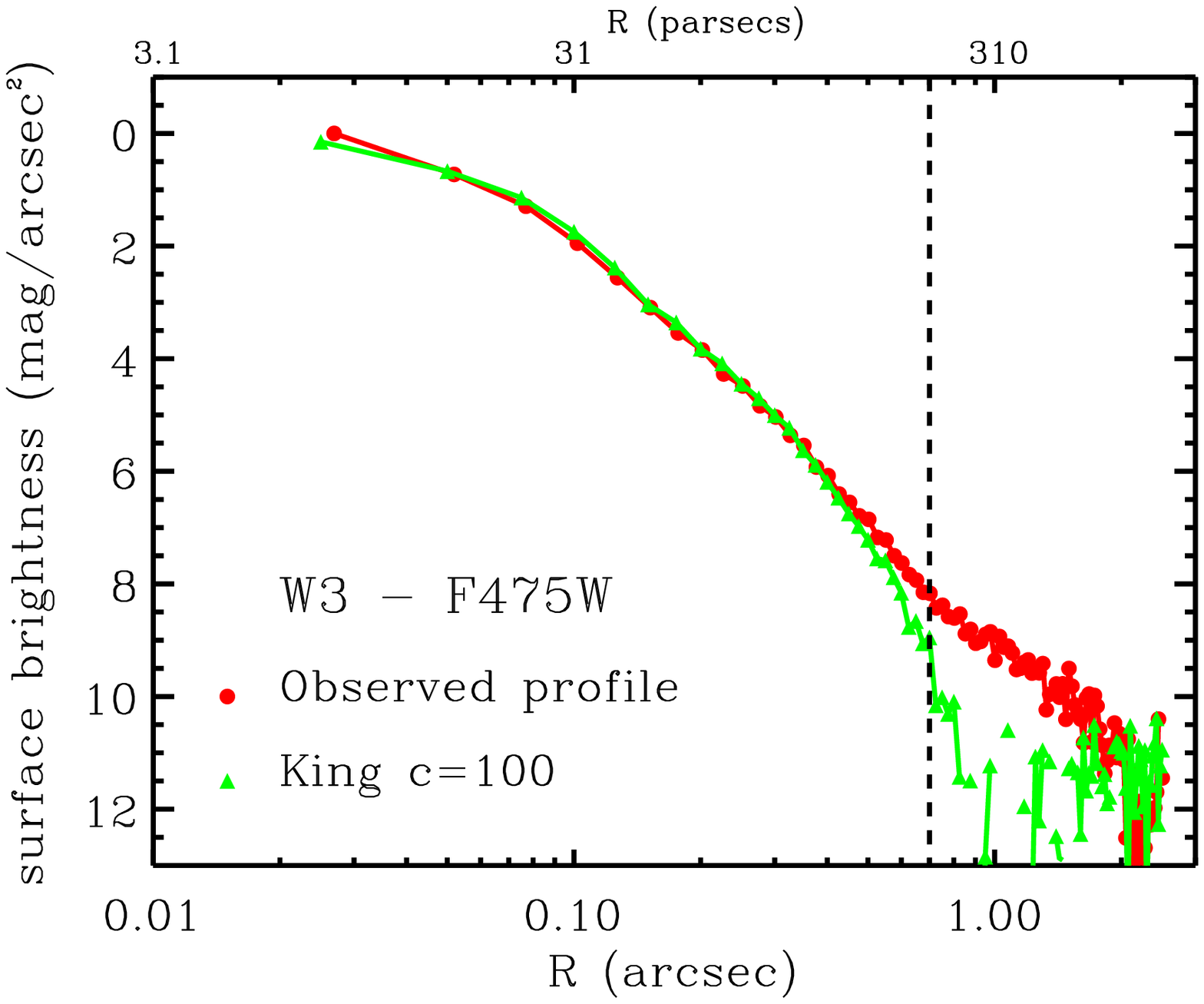}
\includegraphics[width=8.5cm]{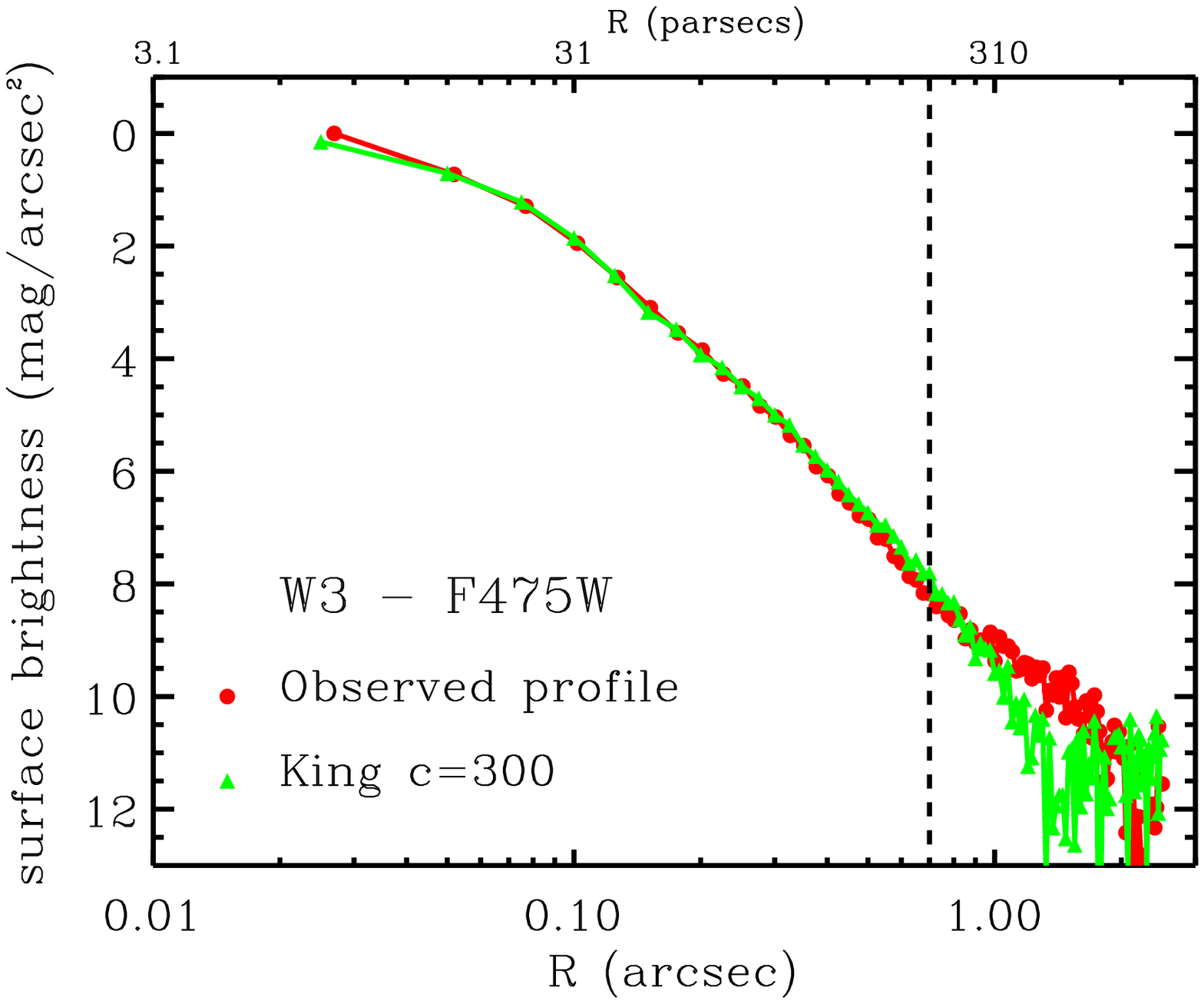}
\caption{The surface brightness profile of W3 (red circles) along with the best fitting King profile (green triangles)  with $c=100$ (top) and $c=300$ (bottom).  The fits were carried out within 0.56\arcsec\ and the model profiles were measured on artificial clusters added to the image at the same galactocentric distance as W3.  The dashed vertical line marks a 0.7\arcsec\ radius ($\sim215$~pc) for reference.  Note that a truncation in the profile in either case would be observable.  If the underlying profile is best described by a King profile, then in order for the truncation to not be seen the concentration parameter would need to be $c>500$.}
\label{fig:w3_profile_king}
\end{figure}

\begin{figure}
\includegraphics[width=8.5cm]{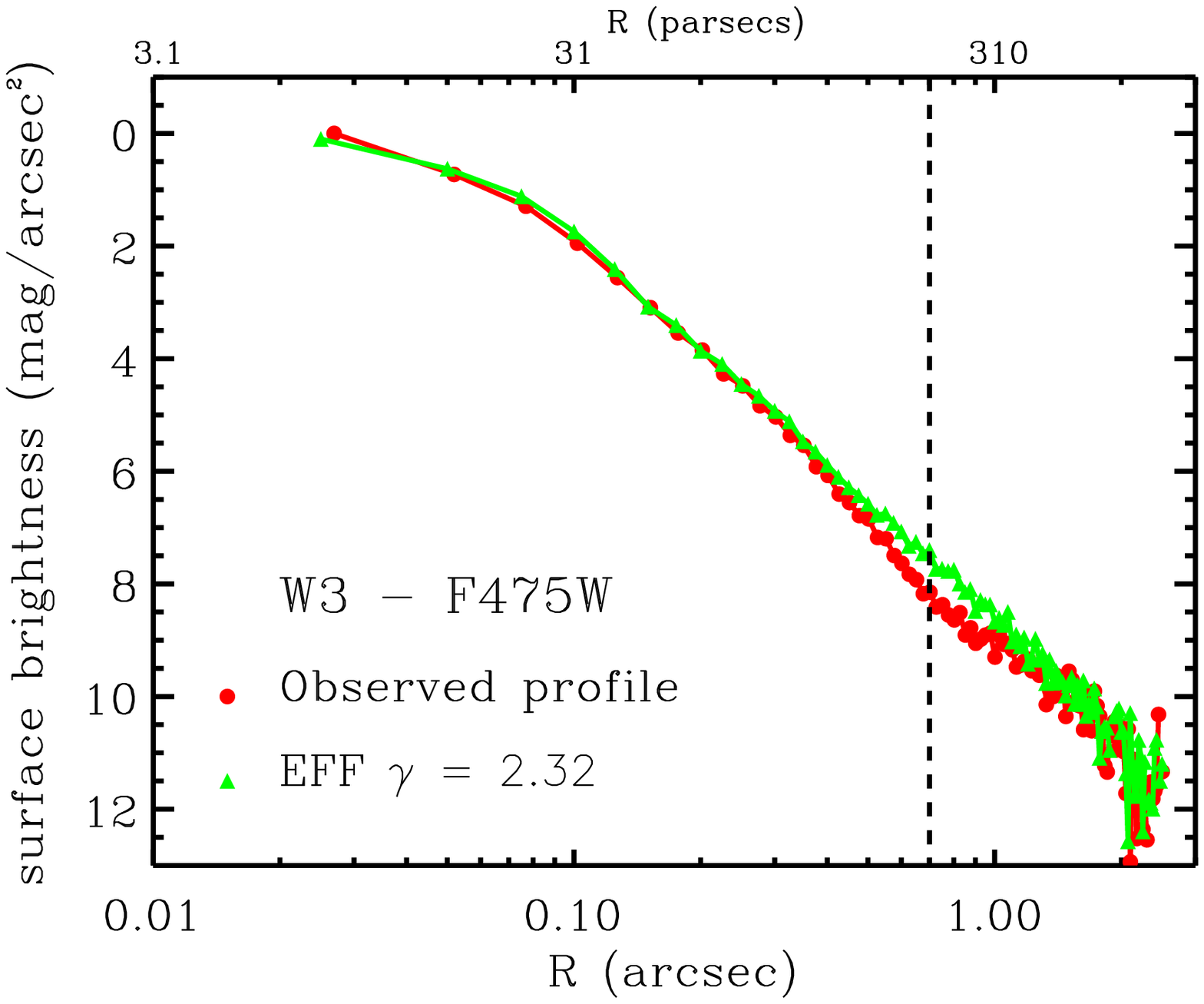}
\includegraphics[width=8.5cm]{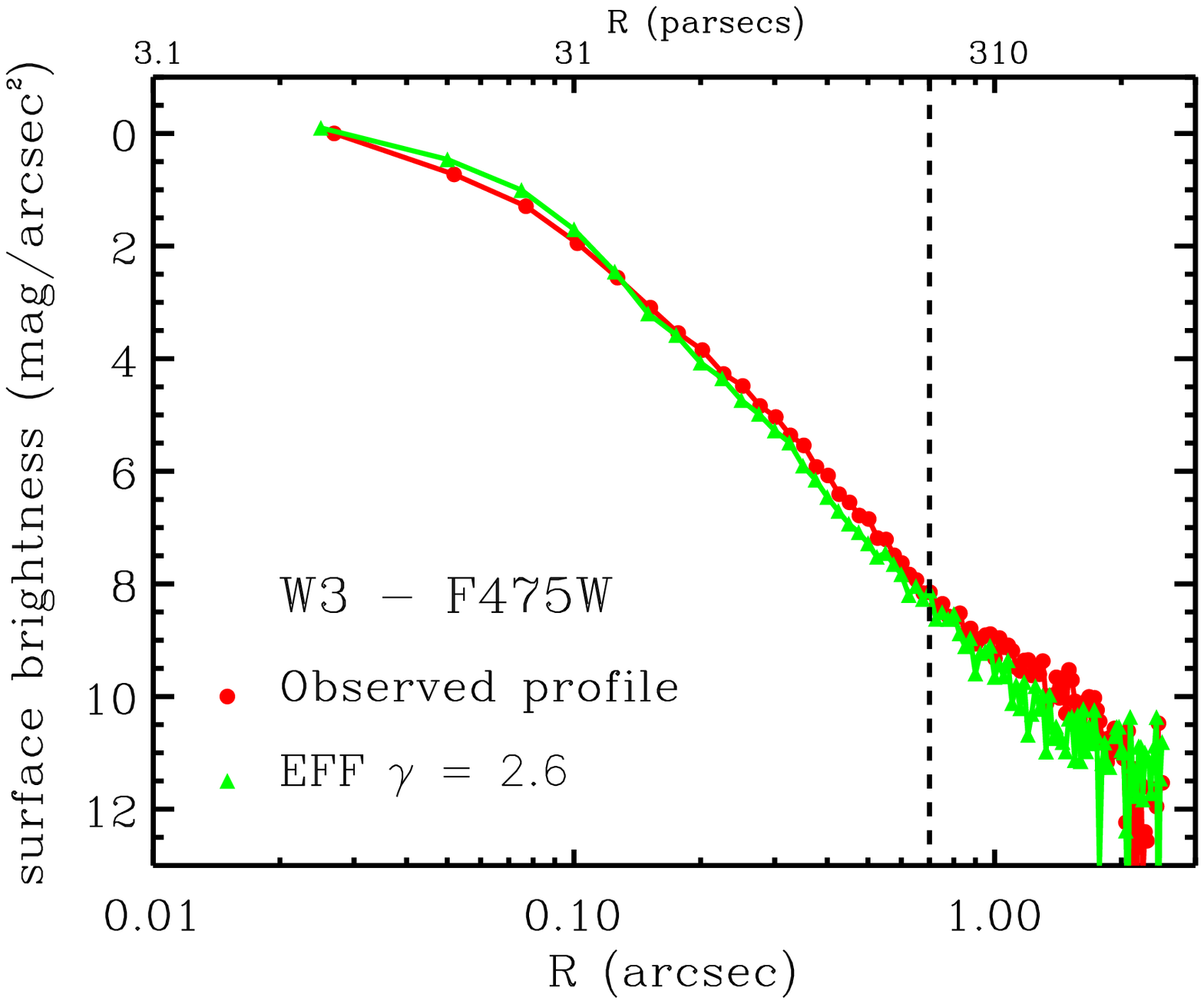}
\caption{Similar to Fig.~\ref{fig:w3_profile_king}, but now the model profiles are for clusters with an EFF profile.  The top panel shows the best fitting (lowest $\chi^2$) profile, while the bottom panel shows a profile with $\gamma=2.6$.  Such power-law like profiles provide a better fit to the outer envelope of W3 than King profiles.}
\label{fig:w3_profile_eff}
\end{figure}

Once the best fitting profiles were found for each profile type, we created artificial clusters of that profile type and radius, with the same brightness as W3 (within 40 pixels) and added them to the images (7 artificial clusters for each profile type) at the same galactocentric distance as W3, but at different position angles.  We then directly compared the SBPs of the models with W3.  Example profile comparisons are presented in Figs.~\ref{fig:w3_profile_king} and \ref{fig:w3_profile_eff} for King and EFF profiles, respectively.

As can be seen from Fig.~\ref{fig:w3_profile_king}, {\em the profile of W3 does not show any signs of truncation} out to where the background was taken at 2.2\arcsec\ ($\approx700$~pc).  Comparing with the best fitting King model profiles of $c=100$ (top panel) and $c=300$ (bottom panel), it is clear that---if present--such a truncation would have been observed.  If we allow $c$ to vary, values of $\sim500$ are preferred.

In contrast, EFF model profiles, which do not have any truncation, provide excellent fits to the data.  Figure~\ref{fig:w3_profile_eff} shows two examples: the lowest-$\chi^2$ profile (within a fitting radius of 20 pixels~= 0.56\arcsec~$\approx$ 175~pc) is shown in the top panel, while a similar profile with $\gamma=2.6$ is shown in the bottom panel.  Both model profiles reproduce the data quite well, and we adopt a best-fit index of $\gamma=2.4\pm0.2$ for the full cluster profile.

The corresponding effective radius is $\reff=17.2_{-2}^{+6}$~pc for both the $F475W$ and $F775W$ images.  The reason for the larger error on the high end is that for shallower indices (approaching $\gamma=2$) \reff\ becomes undefined.  Hence, values of $\gamma$ near 2 lead to large uncertainties in the determination of \reff.

\subsubsection{Dynamical mass}

Using the best-fit model profile (EFF with $\gamma=2.4$) we estimate an effective radius of $17.2_{-2}^{+6}$~pc for W3.  This value is very similar to that found by M04, namely $\reff=17.5\pm1.8$~pc, based on lower-resolution WFPC2 observations.  Hence, we confirm the dynamical mass estimate of $8\pm2\times10^7$\,\msun\ by M04, making W3 the most massive star cluster presently known.



\subsection{W30}
\label{subsec:W30}

The next most massive cluster in the NGC~7252 population is W30, with a measured dynamical mass of $1.6\pm0.3 \times 10^7$\,\msun\ (B06).  W30 is $\sim1.6$~mag fainter in the $V$-band than W3, resulting in a lower S/N, especially in the outer regions.  Additionally, the analysis is complicated somewhat by the presence of two neighbouring sources at $0\farcs92$ and $1\farcs04$.  However, these sources are significantly fainter than W30 on both the $F475W$ and $F775W$ images and can, therefore, be masked out by the {\em ISHAPE} software during the fitting.

The measured SBP of W30 is shown in Fig.~\ref{fig:compare_profiles} as solid (green) circles.  One can readily see by eye that W30 is more compact than W3, although it, too, features an extended profile.  This profile can be traced out to at least 250~pc (0.8\arcsec) from the cluster centre.

Carrying out an analysis similar to that described above for W3, we estimate values of $\reff=7.45\pm0.22$~pc and $10.8\pm1.4$~pc for W30 on the $F475W$ and $F775W$ images, respectively.  The straight average of these two values is $\reff = 9.1 \pm 1.7$~pc, in excellent agreement with the value of $9.3\pm 1.7$~pc found by B06.

The index of the best-fit EFF profile for W30 appears to be similar to that for W3, with $\gamma=2.50\pm0.04$ and $\gamma=2.36\pm0.06$ on the $F475W$ and $F775W$ images, respectively.  There is some indication that W30 is slightly elongated (minor/major axis = 0.84) on the blue image, but no trace of this is found on the red image.

\subsection{W26}
\label{subsec:W26}

\begin{figure}
\includegraphics[width=8.5cm]{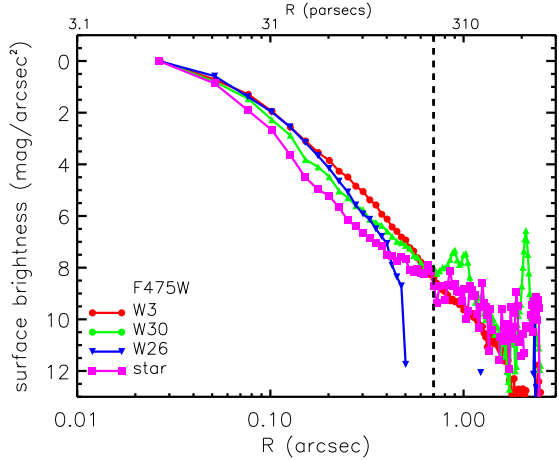}
\caption{Intercomparison of the profiles of three extended clusters, shown along with the profile of a foreground star in the same field of view.  Note that W3 and W30 extend to at least $500$~pc and $250$~pc, respectively.  The fact that the foreground star used as a comparison crosses the extended profiles of W3 and W30 is due to its faintness, making it difficult to reliably trace its profile past 8~magnitudes below its central surface brightness.}
\label{fig:compare_profiles}
\end{figure}

Another cluster that stands out in Fig.~\ref{fig:comp} is W26, the profile of which is shown in Fig.~\ref{fig:compare_profiles}.  This cluster is $\sim0.9$ magnitudes fainter than W30 (M97), making it more difficult to measure the outer parts of its profile.  This may perhaps explain the relatively abrupt truncation observed at $0\farcs5$ ($\sim150$~pc), but alternatively the truncation could also be real (\S~\ref{subsec:tidal_radii}).

We measure $\reff=11.4$~pc and $13.85$~pc in the blue and red bands, respectively, which yields a straight average of $\reff=12.6\pm1.7$~pc.  The EFF-profile index of W26 appears to be steeper than that of W3 or W30, with $\gamma=2.9$ and $2.5$ in the blue and red bands, respectively.  No evidence for any elongation is seen for W26 in either band.

\begin{table} 
\caption{Cluster ID, effective radius, and best fit profile index (assuming an EFF profile) for the clusters discussed in \S~\ref{sec:profiles}.}
  \begin{tabular}
    {lclcccc}
    \hline
 ID  && \phn\phn R$_{\rm eff}$  &$\gamma$ (EFF profile)& Extent (pc)\\
    \hline 
 W3  &&    17.2$^{+6}_{-2}$    & 2.4$^{+0.2}_{-0.2}$ & $>500$\\
 W30 && \phn9.1$^{+2.3}_{-2.3}$& 2.4$^{+0.2}_{-0.2}$ & $>250$\\
 W26 &&    12.6$^{+1.7}_{-1.7}$& 2.7$^{+0.2}_{-0.2}$ & $150$\\
    \hline 
  \end{tabular}
\label{tab:objects}
\end{table}

\begin{table*}
\caption{Magnitudes, concentration indices CI, effective radii, and projected galactocentric distances for 36 young massive clusters of NGC~7252 discussed in \S~\ref{sec:profiles}. Note that all structural parameters were measured assuming an EFF profile with $\gamma=3.0$, which is why some of the values differ from those in Table~\ref{tab:objects}. The observed magnitudes have not been corrected for foreground or internal extinction.  The IDs beginning with 'W' refer to Whitmore et al.~(1993), 'S' refers to S98 and 'Mi' to M97.  For clusters not previously catalogued we provide the coordinates in the final two columns.}
  \begin{tabular}
    {lcccccccc}
    \hline
ID& F336W & F475W & F775W &  CI  & \reff\,(F475W)& \reff\,(F775W)& Galactocentric & RA and DEC  \\
  & (mag) & (mag) & (mag) & (mag)&     (pc)      &     (pc)      & distance (kpc) & (J2000)\\
    \hline 

W3   & 18.70  &  18.21  &  17.38  &   1.64  & 16.7  & 17.7  &   4.70    \\
W6   & 20.45  &  19.94  &  19.09  &   1.54  &  5.1  &  5.1  &   5.59    \\
W26  & 21.36  &  20.80  &  19.86  &   1.70  & 11.4  & 11.9  &   3.25    \\
W30  & 20.26  &  19.81  &  18.98  &   1.51  &  7.5  &  8.3  &   3.37    \\
W31  & 21.75  &  21.51  &  20.74  &   1.52  &  3.2  &  $<2.0$  &   3.46    \\
S105 & 22.05  &  21.62  &  20.82  &   1.50  &  9.2  &  9.5  &  10.48    \\
S114 & 21.98  &  21.57  &  20.75  &   1.50  &  5.5  &  5.4  &   8.79    \\
  --   & 23.31  &  22.82  &  22.00  &   1.54  &  4.4  &  4.8  &   8.82  & 22:20:45.69 -24:41:07.1  \\
W22    & 22.38  &  21.91  &  21.01  &   1.57  &  6.4  &  5.9  &   4.38    \\
W34     & 23.69  &  23.47  &  22.72  &   1.51  &  2.5  &  2.4  &   4.97    \\
 --    & 24.34  &  23.77  &  22.94  &   1.60  &  9.1  &  9.4  &   7.41   & 22:20:43.11 -24:40:48.8 \\
 --    & 24.36  &  23.85  &  23.05  &   1.66  &  9.7  & 13.1  &  11.42  & 22:20:43.52  -24:41:14.1  \\
 W8    & 23.97  &  23.44  &  22.62  &   1.54  &  2.6  &  3.6  &   3.87    \\
W10  & 20.87  &  20.39  &  19.51  &   1.62  &  7.3  &  4.7  &   2.28    \\
 Mi6    & 23.78  &  23.36  &  22.41  &   1.60  &  6.1  &  5.6  &   3.42    \\
  Mi17   & 23.41  &  23.02  &  22.20  &   1.53  &  5.7  &  3.6  &   3.06    \\
  --   & 22.26  &  21.74  &  20.85  &   1.49  &  3.1  &  3.1  &   6.16    & 22:20:43.82 -24:40:26.9\\
W9   & 23.08  &  22.73  &  21.86  &   1.55  &  5.3  &  4.4  &   2.64    \\
W19  & 23.33  &  22.68  &  21.65  &   1.54  &  2.8  &  5.3  &   1.85    \\
W24  & 23.49  &  23.07  &  22.18  &   1.55  &  5.6  &  4.0  &   2.56    \\
W17     & 24.41  &  23.80  &  23.06  &   1.61  &  5.6  &  8.0  &   2.10    \\
 Mi12    & 23.95  &  23.53  &  22.62  &   1.57  & 10.0  & 12.7  &   1.99    \\
  --   & 24.57  &  24.17  &  23.36  &   1.66  & 12.7  & 12.5  &   3.75   & 22:20:43.91 -24:40:39.6 \\
W5   & 23.42  &  23.23  &  22.43  &   1.61  &  9.8  & 10.7  &   3.84    \\
Mi31  & 24.84  &  24.07  &  23.23  &   1.47  &  4.8  &  4.7  &   2.92    \\
--     & 24.83  &  24.25  &  23.52  &   1.61  &  6.6  &  4.8  &   3.49    & 22:20:44.69 -24:40:30.6\\
Mi49     & 24.32  &  23.96  &  23.21  &   1.47  &  4.8  &  5.1  &   1.66    \\
 W20    & 23.87  &  23.21  &  22.42  &   1.53  &  3.5  &  2.2  &   3.82    \\
  Mi35   & 24.24  &  24.07  &  23.18  &   1.60  &  5.3  &  9.9  &   4.12    \\
W28    & 24.09  &  23.44  &  22.40  &   1.65  &  8.7  & 10.4  &   3.09    \\
  --   & 24.18  &  23.73  &  22.88  &   1.48  &  5.7  &  4.4  &   5.53   & 22:20:43.68 -24:40:32.3 \\
  --   & 23.77  &  23.27  &  22.52  &   1.61  &  7.7  &  7.0  &   6.22   & 22:20:43.36 -24:40:36.7 \\
  --   & 24.11  &  23.50  &  22.69  &   1.49  &  4.3  &  3.6  &   9.05   & 22:20:42.80 -24:40:30.9\\
  --   & 24.69  &  24.07  &  23.23  &   1.67  &  9.0  &  9.4  &   5.52   &  22:20:43.51 -24:40:45.3\\
  --   & 25.03  &  24.65  &  23.80  &   1.60  &  6.5  &  7.7  &   4.08   &  22:20:43.82 -24:40:41.1 \\
  --   & 25.03  &  24.80  &  24.07  &   1.68  & 13.1  & 16.9  &  13.21 &  22:20:45.66 -24:41:22.3 \\                                        
    \hline 
  \end{tabular}
\label{tab:catalogue}
\end{table*}

\subsection{Other massive clusters}
\label{subsec:other_YMCs}

In addition to the three clusters with extended profiles discussed above, we attempted to derive the profile shapes for a number of other bright massive clusters in NGC 7252.  Due to the lower S/N of these clusters we did not create artificial cluster models and add them to the image to investigate their profiles directly; instead, we simply used {\em ISHAPE} to estimate $\gamma$, adopting an EFF profile.  In order to test the reliability of the estimated parameters, we ran the fits with three initial guesses for $\gamma$, namely $\gamma = 2.0$, 4.0 and 6.0, and required the estimated values to converge to the same value for all initial guesses.  

In this way, we were able to estimate $\gamma$ values of 2.8, 2.6, 3.0 and 3.0 for W6, S105, S114 and W10, respectively.  These values are near to, although slightly steeper than, those derived for W3, W30 and W26.  Due to the similarity between the estimated $\gamma$ values and that of $\gamma=3.0$ adopted for estimating \reff\ for the full sample, the radii of these clusters do not differ significantly from those given in Table~\ref{tab:catalogue}.



\subsection{Tidal radii}
\label{subsec:tidal_radii}

Given the extended envelopes around W3, W30 and W26, it is insightful to relate their extent to an estimate of the tidal radius for each cluster.  The tidal radius can be estimated via 
$$r_{t} = \left(\frac{G M_{\rm cl}}{2 \cdot V_{\rm G}^2}\right)^{1/3} {\rm R_{\rm G}^{2/3}} $$
where $M_{\rm cl}$ is the mass of the cluster, $R_{\rm G}$ the true distance from the galaxy centre, and $V_{\rm G}$ the circular velocity within the galaxy at that distance (von Hoerner~1957).  This estimate assumes that the clusters are on circular orbits,  and we approximate the true distance with the projected distance.  If the clusters are located further away from the galaxy centre their tidal radii would be larger, and if the clusters move on radial or elliptical orbits (i.e., passing near the centre of the galaxy) their tidal radii would be significantly smaller than estimated via the above equation and approximations.  Additionally, the above approximation may not be accurate if the three extended clusters are located at a true galactocentric distance near their projected one.  In such a case, each cluster's extent would be a significant portion of the galactocentric distance, violating the initial assumptions of the approximation. 

For the masses of W3, W30 and W26 we use $8\times10^7$\,\msun\ (M04), $1.6\times10^7$\,\msun\ (B06) and $6\times10^6$\,\msun\ (assuming the same age as W3 and W30), respectively.  Additionally, we assume a circular velocity of $\sim200$~\kms.  With these values we estimate the tidal radii of W3, W30 and W26 to be approximately $450$~pc, $200$~pc and $150$~pc, respectively.   Given the various uncertainties associated with these estimates, the observed extents of the profiles of W3, W30 and W26 given in Table~1 appear largely consistent with the estimated tidal radii.
Specifically, the observed extents of the envelopes of W3 and W30 are at least equal to the estimated tidal radii and may exceed them, suggesting that either the true galactocentric distances of these two clusters exceed the projected distances significantly or there has been insufficient time for significant tidal erosion.  The near agreement between the measured extent of W26 and the estimated tidal radius, on the other hand, suggests that the observed truncation may be tidal and real, rather than due to a measuring problem.  Our main conclusion is that the observed large extents and/or tidal radii of these clusters are due, at least in part, to the high cluster masses.  Hence, extremely extended profiles may be a common feature of very high-mass clusters.

\begin{figure}
\includegraphics[width=7.5cm]{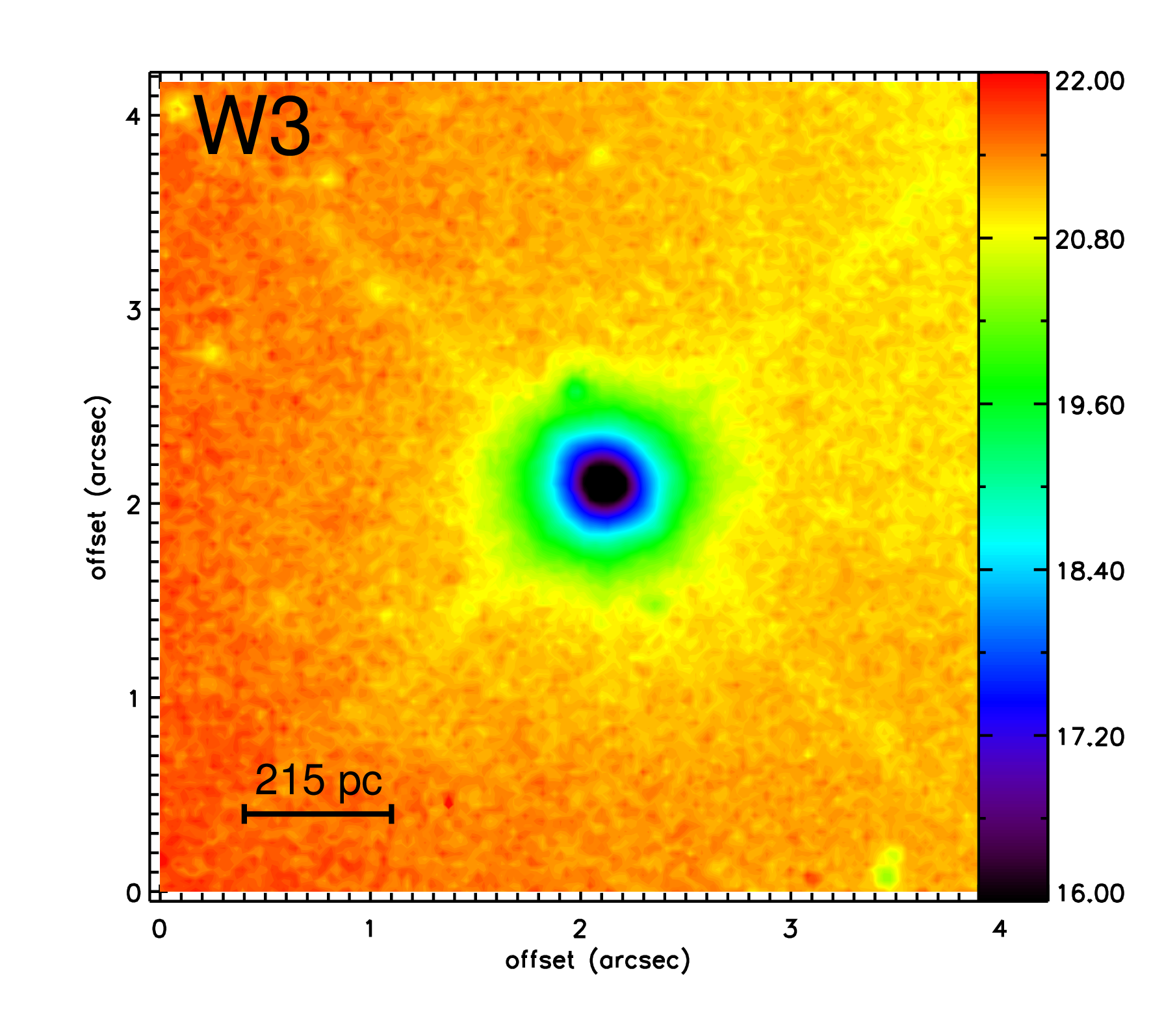}
\includegraphics[width=7.5cm]{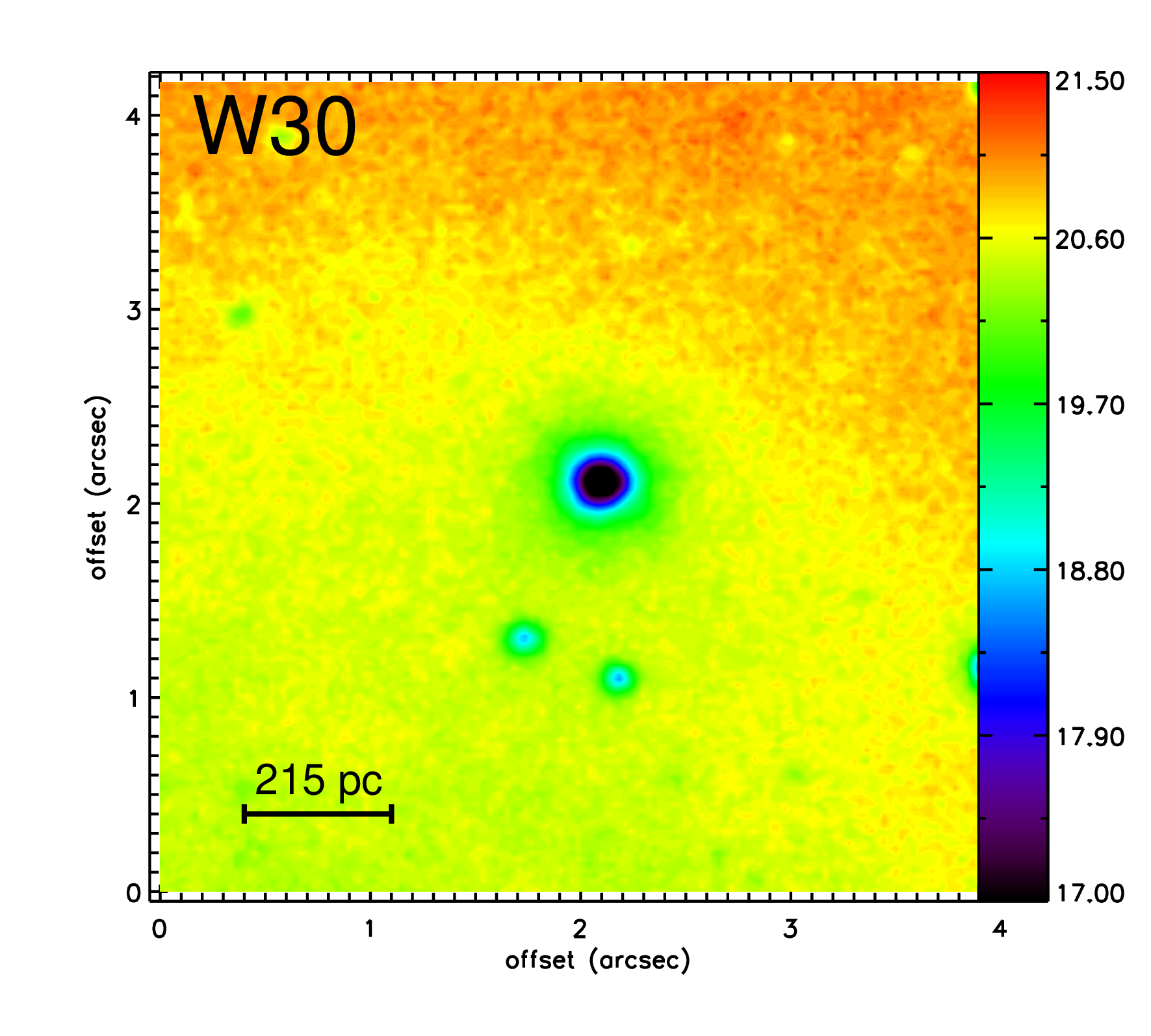}
\includegraphics[width=7.5cm]{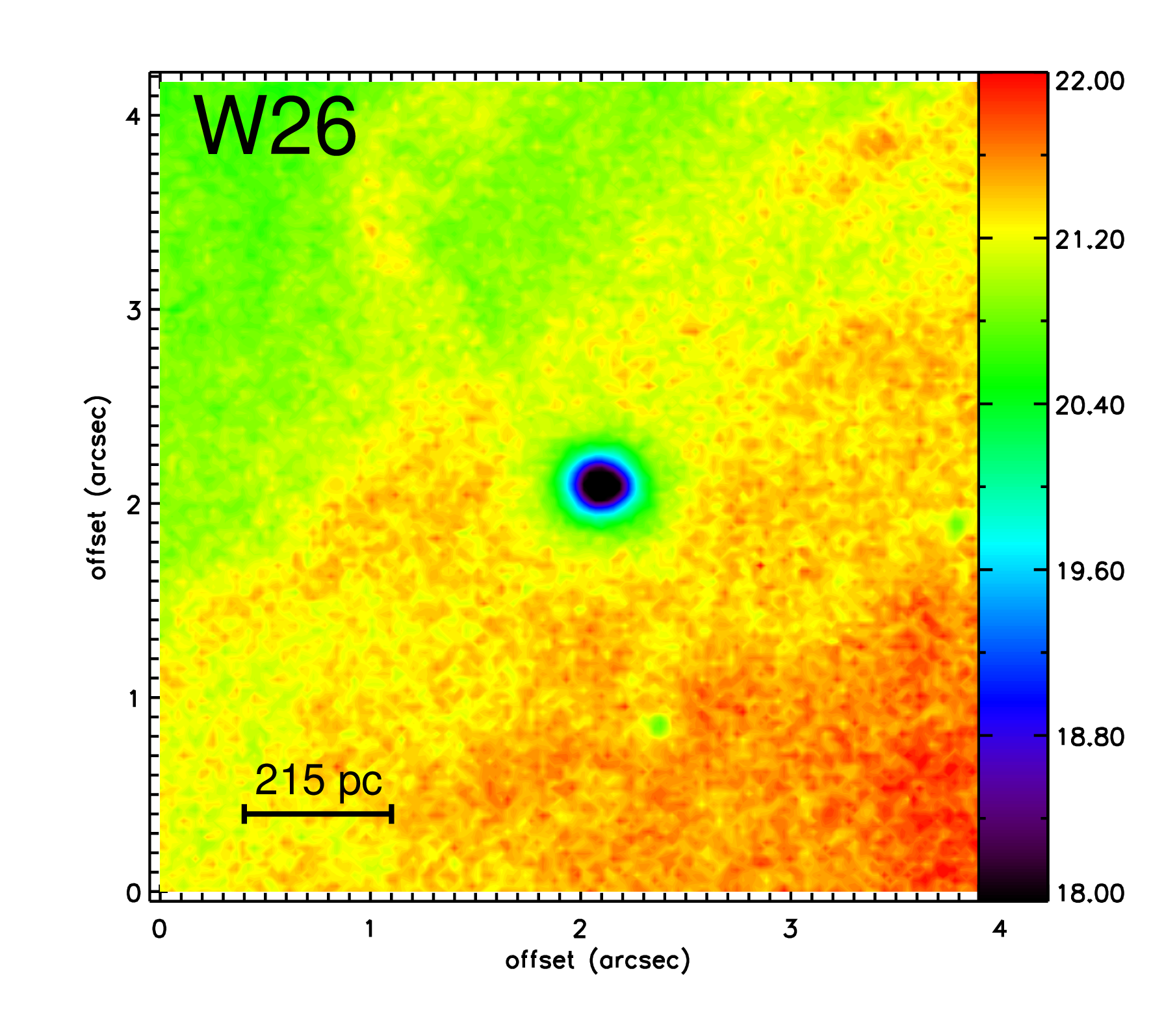}
\caption{Surface brightness maps (F475W) of the three clusters with profiles that extend beyond 150~pc.  Each panel is 4\arcsec\ on a side (1.2~kpc) and no smoothing has been applied. The (relatively smooth) gradient in the background stellar brightness (from NGC~7252) can be seen in each panel.  In the inner region of the clusters, each pixel has been set to the minimum surface brightness shown in the colour bar (i.e. the inner sections are not resolved in this representation).  Note that all three clusters appear circular in projection and that no tidal features are apparent.  However, due to the relatively bright background, any possible tidal features would be difficult to discern.}
\label{fig:sb_maps}
\end{figure}

Due to the large extent of W3, W30 and W26, we have also searched for tidal debris around each cluster.  Figure~\ref{fig:sb_maps} shows surface brightness maps (mag/arcsec$^2$) in the F475W image, of each cluster.  All three clusters appear circular and no clear tidal debris is seen.  From these images we can see that a the weakly varying background is not affecting our results (e.g., spiral structure, dust lanes, etc) and that binning the profile of the clusters radially (i.e., averaging azimuthally) does not hide structure.  Tidal features, while not visible in the surface brightness images, may still be present; however the relatively bright background limits the possibility of detecting such structures with the present data.

\section{Radius distribution and the mass--radius relation}
\label{sec:radii}

While we cannot fit SBPs for the other cluster candidates in our sample due to their lower S/N and blending with the background galaxy light, we have attempted to measure \reff\ for each of the candidates.   For this we only use sources that are found to be resolved with both {\it ISHAPE}\, and the concentration index technique (see Fig.~\ref{fig:comp}), i.e. 36 of the 52 cluster candidates identified through their colours (see \S~\ref{sec:sample}).  

We find that both the mean and median values of \reff\ are $\sim$~6\,--\,7 pc, independent of the filter used.  This is significantly larger than for typical YMCs ($\sim2.5$~pc; e.g., Larsen~2004; Portegies Zwart et al.~2010).  There are three possible explanations for this difference.  The first is that the difference is real and that our subsample of clusters in NGC~7252 is representative of the size distribution of the cluster population as a whole within this galaxy.  The second possibility is that we have biased our sample by only considering well resolved clusters.  This is certainly affecting our sample at some level: for example, if we assign all the unresolved clusters an effective radius of 2~pc, the median of the distribution is $\lesssim5$~pc.  Finally, it has been suggested that there is an underlying mass--radius relation for clusters above $10^6$\,\msun\ (e.g., Ha{\c s}egan et al.~2005; Kissler-Patig, Jord{\'a}n, \& Bastian~2006).   Many of the clusters in our sample have masses above $10^6$\,\msun, so their large sizes may simply be a reflection of their high masses.

\begin{figure}
\includegraphics[width=8.5cm]{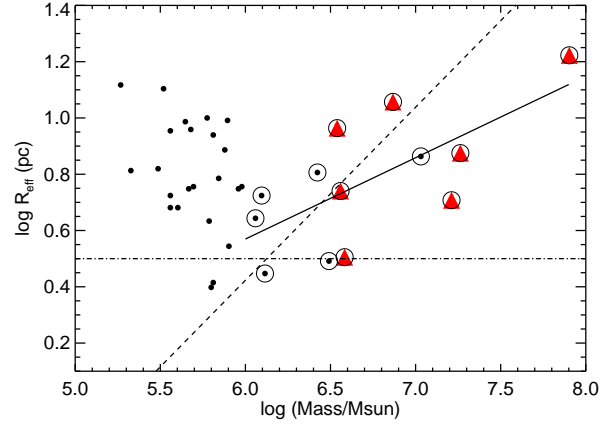}
\caption{The relation between mass and effective radius for the young massive clusters of NGC~7252.  Masses were determined by assuming that all clusters in the sample have the same age (due to our colour selection), suffer no extinction, and can simply be scaled from their $F475W$ magnitudes.  The (red) triangles mark spectroscopically confirmed clusters, while the open circles denote clusters that were used in the regression fit.  The best-fit relation is shown as the solid line, while the dashed line shows the relation for luminous elliptical galaxies (Ha{\c s}egan et al.~2005) for comparison, and the dash-dotted line marks the mean effective radius of 3.2~pc for old globular clusters. For masses above $1\times10^6$\,\msun\ we find evidence for a mass--radius relation at the $\sim3\sigma$ level.}
\label{fig:mass-radius}
\end{figure}

In order to test this third possibility we searched for a mass--luminosity relation of the resolved cluster candidates that pass our colour selection.  First, we translated each cluster's $F475W$ magnitude to a mass by assuming that all clusters have the same age ($400$~Myr)---due to our colour selection---and solar metallicity (S98) and that they are not affected significantly by extinction.  The result is shown in Fig.~\ref{fig:mass-radius}.  The dashed line represents the extrapolated relation from luminous elliptical galaxies (see Ha{\c s}egan et al.~2005), while the dash-dotted horizontal line marks the mean globular cluster radius, which is similar to our resolution limit.  We performed a regression fit to the data above $1.0\times10^6$\,\msun\ (indicated by open circles) of the form:\, $\log (\reff/{\rm pc}) = a + b*\log(M_{\rm cl}/\msun)$, where $M_{\rm cl}$ is the cluster mass, and found $a=-1.16\pm0.63$ and $b=0.29\pm0.09$; this fit is shown as a solid line.  From this fit we find {\em evidence for a mass--radius relation for clusters in NGC~7252 with masses above $1.0\times10^6$\,\msun} at the $\sim3\sigma$ level.  However, the exact significance of the correlation depends on the adopted limits (it is lower if a lower mass limit of $2.5\times10^6$\,\msun\ is used).  Additionally, if W3 is excluded from the fit, the significance of the mass--radius relation drops to $\sim1.5\sigma$.

\section{The effective radius -- galactocentric distance relation}
\label{sec:reff_dgc}

\begin{figure}
\includegraphics[width=8.5cm]{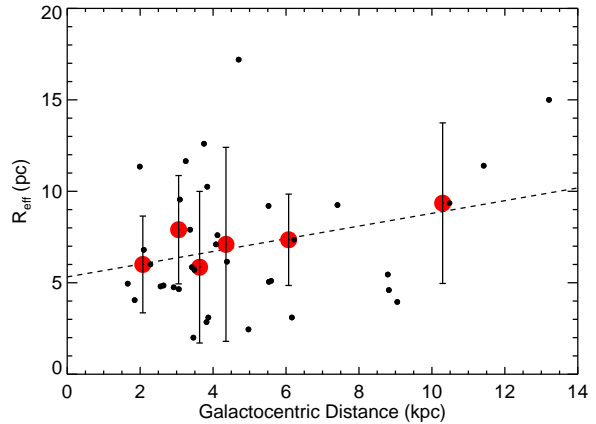}
\caption{The observed relation between \reff\ and projected galactocentric distance for the 36 resolved clusters from Table~\ref{tab:catalogue}.  The small filled circles represent the individual clusters,  while the large (red) filled circles with error bars represent the mean \reff\ and standard deviations of the clusters in distance bins containing six clusters each.  The dashed line shows the linear-regression fit to the unbinned data.}
\label{fig:reff_dgc}
\end{figure}

 Figure~\ref{fig:reff_dgc} shows the measured effective radii of the resolved clusters in our sample plotted versus the projected galactocentric distances $D_{\rm GC}$.  The radii were measured from the $F475W$ image, using an EFF $\gamma=3.0$ model profile, and are given in Table~\ref{tab:catalogue}.  Each cluster is represented by a small filled circle, while the large (red) points with error bars represent the mean values and standard deviations of \reff\ determined in six distance bins with six clusters each.  The dashed line shows a linear regression fit to the individual clusters, which has a slope of $0.35\pm0.20$.

In order to check the validity of the observed \reff\,--\,$D_{\rm GC}$ relation we carried out a series of Monte Carlo simulations.   First, we created a population of 10\,000 clusters with a true spatial distribution made to match the observed projected distribution of clusters in NGC~7252 (M97, esp.\ Fig.~19).  Next we assigned a {\em random} \reff\ to each cluster, chosen from a gaussian distribution with a mean of 7~pc and a standard deviation of 4~pc.  We then selected 36 clusters from the original 10\,000 at random, matching their parameters to those of the observed population ($1.85~{\rm kpc} < D_{\rm GC} < 13.21~ {\rm kpc}$, $\reff>2$~pc).  Finally, we carried out a linear regression on these clusters in the same way as was done for the observations.  This type of analysis implicitly assumes that the clusters move on circular orbits.

\begin{figure}
\includegraphics[width=8.5cm]{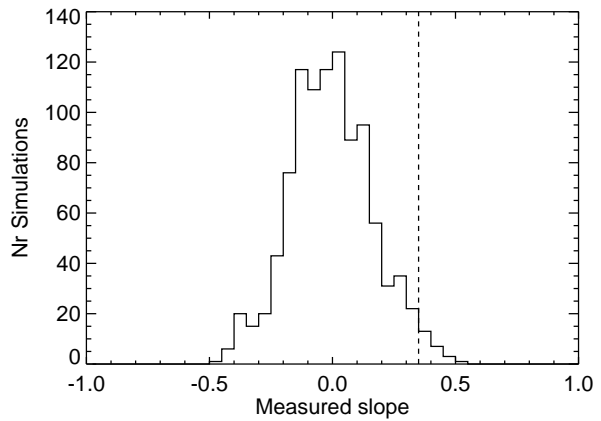}
\caption{The distribution of measured slopes from 1000 Monte Carlo simulations of the NGC~7252 cluster population, assuming no intrinsic relation between \reff\ and galactocentric distance.  The vertical dashed line marks the observed slope  (see Fig.~\ref{fig:reff_dgc}).  Only approximately 5\% of the simulations yield regression slopes equal to or larger than the observed slope.}
\label{fig:sims}
\end{figure}

One thousand realisations of the model were performed.  Figure~\ref{fig:sims} shows the resulting distribution of the slopes of linear regressions performed between \reff\ and galactocentric distance.  The vertical dashed line marks the observed slope (shown in Fig.~\ref{fig:reff_dgc}).  We find that only $\sim5$\% of the random simulations resulted in a slope similar to or larger than that observed.  

We conclude that there is an intriguing relation between \reff\ and the galactocentric distance of clusters in NGC~7252.  However, the observed relation {\em may} be due to small-number statistics.  Expanding the present study to other galactic merger remnants with massive clusters may allow a more definitive answer.  Goudfrooij~(2012) looked at the \reff\,--\,$D_{\rm GC}$ relation for massive clusters in the intermediate age ($\sim3$~Gyr - e.g., Goudfrooij et al.~2001) merger remnant NGC 1316.  He found a similar relation as that reported here, that the mean effective radius increases as a function of projected galactocentric distance.  This lends support to the notion that such a relation may be intrinsic to young-to-intermediate-age cluster populations in merger remnants.


\section{Discussion}
\label{sec:discussion}

\subsection{Comparison with Knot S in the Antennae Galaxies}
\label{subsec:comparison_wKnotS}

While W3 and W30 are two of the most massive star clusters known, they are not unique in having extended power-law envelopes.  The Antennae galaxies host a number of massive clusters as well as extended ``knots'' of ongoing or recent star and cluster formation.  The brightest such knot, known as `Knot S' (Rubin et al.\ 1970; Whitmore et al.\ 1999), is located in the distorted disk of NGC~4038, away from the ``overlap region'' where the two galaxies' gas masses are currently colliding; see Fig.~28 in Whitmore et al. (2010) for an {\em HST}/ACS-HRC image of this knot.

\begin{figure}
\includegraphics[width=8.5cm]{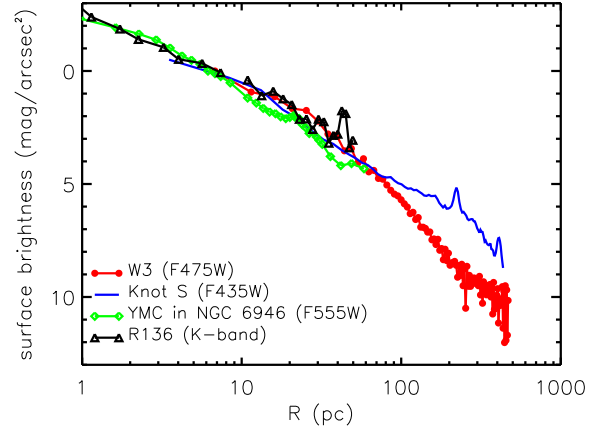}
\caption{Comparison of the surface-brightness profile of Knot S in the Antennae galaxies and a YMC in NGC~6946 with that of W3.  Knot S also has a power-law profile that extends to at least 400~pc, although its profile is shallower than that of W3.  This difference could be intrinsic (i.e. present at formation) or alternatively due to evolution as W3 is $\sim400$~Myr old, while Knot S is $\sim7$~Myr old.  A YMC in NGC~6946 is also shown (green) for comparison.  This cluster also has an extended power-law envelope, as discussed in Larsen et al.~(2001).  Beyond the radius shown for this cluster, the profile becomes flat, due to the presence of a related stellar complex.}
\label{fig:knots}
\end{figure}

We used {\em HST}/ACS-WFC $V$-band images (presented in detail in Bastian et al. 2009; Whitmore et al. 2010) to study the profile of Knot S (following Schweizer 2004).  For the analysis we used the empirical PSFs described in Bastian et al.~(2009).  The profile of Knot S is shown in Fig.~\ref{fig:knots} as a solid (blue) line.  We measured the profile of Knot S in the same way as for the NGC~7252 clusters discussed above, namely using {\em ISHAPE}, selecting an EFF model profile and fitting on the FWHM and index $\gamma$ of the profile.  The initial estimate of \reff, for a fitting radius of 25 ACS pixels = 120 pc, yields $\reff = 18$~pc, however with a shallow profile of $\gamma\approx 2$.   Since the profile is so shallow, the effective radius is not well constrained.  If we fix a profile type and shape (EFF, $\gamma=3.0$) and fit both Knot S and W3 with it, we find that both fits result in $\reff \approx 15$\,--\,20~pc.  Hence we conclude that both clusters have similar characteristic radii.  Both also have envelopes of similar extent ($>$\,450~pc).

An interesting property of Knot S is its youth.  Whitmore et al. (2010) estimate an age of $\sim$\,6~Myr for the core of the knot, i.e. within 15~pc from the centre, through broad-band photometric techniques, and Whitmore et al.~(1999) estimate a similar age, $\sim$\,7~Myr, based on UV spectroscopy.  Additionally, the significant amount of substructure (i.e. small subclusters) present in the halo of Knot S suggests that it is dynamically unevolved, i.e. the current distribution is similar to the initial structure of the knot (although the inner regions may be dynamically more evolved).  The subclusters in the envelope of Knot S show evidence for a range of ages from $\sim1$~Myr to $\sim30$~Myr. The oldest such subcluster is a massive cluster of $\sim1 \times10^6$\,\msun, seen as a bump in the SBP at $R\approx 220$~pc.  It is currently unclear whether this cluster, which has a photometric age of 30~Myr, is associated with the knot, is merely a chance superposition, or is currently just passing near the knot.

Whether or not this massive (sub)cluster is part of Knot S, it is clear that there is a significant age spread amongst the stellar populations present.  This is not surprising, given the large length scales involved.  In nearby galaxies, such as in the LMC, there is a clear correlation between the age difference between two star clusters and the physical distance between them (Efremov \& Elmegreen~1998). These authors interpret this as being due to the spatial extent of the progenitor giant molecular clouds with respect to the sound speed.  Based on these authors' schematic representation (made for the LMC), for a size of $\sim$\,200\,--\,500~pc, we would expect an age spread of 20\,--\,30~Myr, in good agreement with the observations.  Based on this, we may expect W3 to have a similar, or even larger, age spread of stellar populations within it.   

Fellhauer \& Kroupa~(2005) have presented scaled $N$-body models of a complex of massive star clusters.  In their models, the population of clusters is distributed in a centrally concentrated way; i.e. the clusters themselves follow a Plummer sphere with radius of $\sim$\,100~pc.  The stars within each cluster are in virial equilibrium, and the motions and distribution of clusters globally are also in virial equilibrium.  These authors show that (1) the clusters then merge in a relatively short timespan (a few Myr to a few 100~Myr based on simulations scaled to match W3) and (2) the resulting cluster features an extended envelope reaching out to $\sim1$~kpc.

While Knot S clearly has substructure within its halo, one of the predictions of the Fellhauer \& Kroupa~(2005) model is a dynamically cold core plus an extended profile, each of which can be described by a King or EFF profile.  The resulting combined profile features a distinct bump where the two profiles join; i.e. the envelope has more stars than would be predicted from the extension of the inner profile into the outer parts.  No such bump is seen in our profile of Knot S, although this cluster is probably too young to show such a behaviour, nor is such a bump seen in W3.  The reason for this may be Fellhauer \& Kroupa's (2005) assumption that {\em all}\, stars in the complex form in massive clusters which need to merge in order to disrupt.  Possibly an initially smooth distribution of star formation with some clusters in it might dampen the resulting bump in the profile.  Additionally, the resulting profile might differ if the simulations began with a single dominant cluster at the centre of the complex surrounded by smaller clusters, instead of a number of massive clusters spread throughout the initial complex.

\subsection{Extended profiles in other YMCs}
\label{subsec:extended_profiles}

While Knot S is the most massive known potential young counterpart to W3, being $\sim$\,50 times less massive than W3 was at birth (when taking only stellar evolution into account), it is not the only extended cluster known.  Larsen et al.~(2001) studied the mass and profile of a $\sim$\,15~Myr old YMC in the nearby face-on spiral galaxy NGC~6946.  The profile of this cluster, shown as (green) diamonds in Fig.~\ref{fig:knots}, closely resembles that of W3 within the common radial range.
Using {\em HST}/WFPC2 images, Larsen et al.~(2001) traced the profile out to 60~pc and found that it is well described by a power-law profile with a core radius of 1.2~pc and $\gamma=2.1$. 

The dynamical mass of this YMC is $1.7\times10^6$\,\msun\ within 65~pc of the cluster centre.  Beyond this radius the profile becomes flat due to the presence of an underlying stellar association/complex that surrounds the cluster.  It is possible that after this surrounding association dissolves with time, the remnant profile will resemble that of W3 beyond 65~pc.  Like the younger Knot S, this YMC in NGC~6946 also displays substructure in its envelope, with concentrations of stars and also smaller clusters.  Larsen et al.~(2002) find an age spread within the complex of $\sim30$~Myr based on modelling the formation history of resolved stars.

Closer-by, R136, the central star cluster in 30~Doradus, is also known to feature an extended profile (e.g., Moffat et al.~1994).  Recently, Campbell et al.\ (2010) have studied R136 with the {\em VLT} using the  Multi-Conjugate Adaptive Optics Demonstrator (MAD) to obtain images in the $H$ and $K$ passbands.  The area surrounding R136 is rich in gas and dust, and differential reddening effects can introduce spurious structure into the measured profile (e.g., Mackey \& Gilmore~2003; Campbell et al.~2010).  Hence, imaging in the near-IR has the advantage of significantly reducing this bias.  The profile of R136 from the Campbell et al.~(2010) study is shown as open (black) triangles in Fig.~\ref{fig:knots} for radii below 10~pc; it has a best-fit EFF-profile index of $\gamma=1.6$.  For radii larger than 10~pc we use the {\em VLT}/HAWK-I $K$-band profile derived from the integrated light, normalised to match the Campbell et al. profile at 10~pc.  

This $K$-band profile of R136 matches the optical profiles of the YMC in NGC~6946, Knot S in the Antennae and W3 in NGC~7252 out to $\sim$\,50~pc.  Moffat et al.~(1994) report that the same power-law profile continues out to $\sim$\,100~pc (outside the HAWK-I field of view).  Note that R136, along with its power-law envelope, makes up the dominant stellar population of 30~Doradus.  30~Doradus itself, over its full 100 pc scale extent, is known to have an age spread within its stellar population of $\sim$\,10~Myr (Walborn \& Blades~1997).

In contrast, the Galactic cluster NGC~3603, which is similar to R136 in terms of radius, mass, density and age (Portegies Zwart, McMillan, \& Gieles~2010) appears to have a truncated profile, at least for the massive stars, at $\sim$\,1~pc (Moffat et al.~1994).  The youth of both NGC~3603 and R136 suggests that the differences between their profile types reflect different initial conditions rather than different stellar-dynamical evolution (see Ma{\'{\i}}z-Apell{\'a}niz~2001 for a discussion).

\subsection{A census of young massive stellar groupings}
\label{subsec:census}

Beyond these individual clusters, Ma{\'{\i}}z-Apell{\'a}niz~(2001) studied a sample of young ($<$\,20~Myr), relatively massive ($>3\times 10^4$\,\msun) stellar aggregates in nearby galaxies with {\em HST}\, images.  He found that he could broadly split his sample of 27 objects into three categories, with some overlap between them: star clusters with ``weak halos'' (e.g., NGC~3603 types), star clusters with ``strong halos'' (e.g., R136 types), and stellar aggregates that lack a distinct core (e.g., massive OB associations).  Within this categorization, R136, the YMC in NGC~6946, Knot S in the Antennae and W3, W30 and W26 in NGC~7252 would all be classified as clusters with ``strong halos''.  It is worth noting that all of the above galaxies with ``strong halo'' clusters also contain more compact ``weak halo'' clusters.

It then appears that star clusters feature a range of outer profiles, with ``strong halos'' referring to clusters with shallow profiles, and ``weak halos'' to those with steep profiles.

As noted by Ma{\'{\i}}z-Apell{\'a}niz~(2001), the youth of many of the clusters and associations in his sample argues for the profile types to be due to the initial conditions during cluster formation rather than to dynamical processes after formation.  Do the profiles of clusters (extended, truncated, steep or shallow) vary systematically as a function of galactic location and environment?  If so, what is the dominating factor?  A systematic study of YMC profile types in nearby galaxies is required to answer these questions.

\subsection{Extended Envelopes in the NGC~7252 clusters}
\label{subsec:extended_envelopes}

While shallow and extended profiles have been found before for YMCs in a variety of galaxies (see above), the extended envelopes seen in the clusters of NGC~7252 are remarkable for at least two reasons.  The first is their sheer extent.  The envelopes of W26, W30 and W3 extend out to $\sim$\,150~pc, $>$\,250~pc and $>$\,500~pc, respectively.  Only Knot S in the Antennae compares in spatial extent (see Fig.~\ref{fig:knots}).  

The second reason is the remarkable fact that these three clusters have been able to retain their extended envelopes for 300\,--\,500~Myr.  Knot S is only $\sim$\,7~Myr old and hence may lose its extended envelope over the next several 100~Myr.  Similarly, one might expect the $\sim$\,400~Myr old clusters in NGC~7252 to have lost their envelopes due to tidal truncation.  That this did not occur for W3, W30 and W26 is likely due to their extreme masses ($>5\times 10^6$\,\msun).  Interestingly, it is the lowest-mass cluster of these three, W26 ($\sim 6\times 10^6$\,\msun) that
shows possible signs of tidal truncation at $\sim$\,150~pc (Fig.~\ref{fig:compare_profiles} and
\S~\ref{subsec:tidal_radii}). 


\section{Conclusions}
\label{sec:conclusions}

We have presented an analysis of {\em HST}/WFC3 images of the galactic merger remnant NGC~7252.  In particular, we have focussed on the structure and effective radii of the young massive clusters that formed during the merger.  We have been able to resolve clusters down to $\sim2$~pc, despite the 64~Mpc distance to the host galaxy.   The average effective radius of the resolved clusters in our sample is $\langle\reff\rangle\approx 6$\,--\,7~pc, significantly larger than for YMCs in normal galaxies, although we note that this may be caused by the fact that we cannot resolve  the smallest clusters.  We have found some evidence for a relation between the effective radius and mass for clusters with masses $> 1 \times10^6$\msun, and also tentative evidence for a relation between the effective radius and galactocentric distance, with clusters with larger projected distances having larger effective radii.

For three of the brightest clusters, W3, W30 and W26, we have been able to measure surface-brightness profiles and constrain the structural parameters.  We find that all three have power-law envelopes, well fit by EFF profiles
to near the limit of measurability.  These envelopes extend out to $>$\,500~pc, $>$\,250~pc, and $\sim$\,150~pc, for W3, W30, and W26, respectively.  Only the envelope of W26 shows a sign of being truncated (at $\sim$\,150~pc), but whether this apparent truncation is real and tidal, or due to imperfect background subtraction, cannot presently be established with certainty.  We note that it has been possible to trace the profiles out to these large distances mainly because of the brightness (high mass) of these three clusters and the relatively low and smooth background of the host galaxy.

We have compared these three measured profiles to those of other YMCs found in the literature.  A number of young clusters are known to also feature extended profiles, but only one---Knot S in the Antennae galaxies---extends as far out as those of the three NGC~7252 clusters.  It is interesting to compare Knot~S with clusters W3, W30 and W26, due to the youth of its central  region ($\sim7$~Myr).  The envelope of Knot~S contains a number of subclusters that may disrupt over the next few hundred Myr, leaving a smooth halo, as observed in the NGC~7252 clusters.  The subclusters in Knot~S show an age spread of $\sim30$~Myr.  Hence, if the progenitors of W3 and its siblings resemble Knot~S, we may expect similar age spreads within them as well.

The similarities between W3, Knot~S, and other YMCs with extended envelopes suggest that the envelopes and their substructures are primordial in nature.  Other very young clusters, such as NGC~3603 in the Galaxy, appear to form with a distinct truncation.  We speculate that the difference may stem from the gas distributions at the time when star formation begins: Gaseous cores distinct from their surroundings may form truncated star clusters, while core--cloud continua may form clusters with extended envelopes.

In the process of our analysis, we have estimated the tidal radii of the three extended clusters in NGC 7252 from their estimated masses and projected galactocentric distances, assuming that the clusters move on circular orbits.  We find that these tidal radii are comparable to the observed extents of the profiles.  This may explain why these very massive clusters ($> 5\times 10^6$\,\msun) were able to retain their extended envelopes for several hundred Myr.  Other young clusters that display extended envelopes, such as R136 in the LMC or the YMC in NGC~6946, may lose their envelopes more quickly due to their lower masses and more disk-like environments.


\section*{Acknowledgments}

We thank the HAWK-I instrument and commissioning teams for making the 30~Dor frames available and John Pritchard for his advice concerning those images.  The present paper is based on observations taken with the NASA/ESA {\em Hubble Space Telescope}, obtained at the Space Telescope Science Institute, which is operated by AURA, Inc., under NASA contract NAS5-26555.  The research was supported by the DFG cluster of excellence `Origin and Structure of the Universe' (www.universe-cluster.de).  N.B. is funded by a University Research Fellowship from the Royal Society.  F.S. gratefully acknowledges support by NASA through Grant GO-11\,554.01-A from the Space Telescope Science Institute and by the Carnegie Institution for Science.  P.G. was partially supported by NASA through Grant GO-11691.01-A from the Space Telescope Science Institute.  Part of this work was done as part of an International Team at the International Space Science Institute in Bern.

\bsp
\label{lastpage}

\begin{thebibliography}{99}

\bibitem[\protect\citeauthoryear{Anderson \& King}{2006}]{andkin06}
Anderson J., King I. R. 2006, ``PSFs, Photometry, and Astrometry for the
 ACS/WFC'', ACS Instrument Science Report 2006-01 (Baltimore:STScI)

\bibitem[Bastian et 
al.(2005)]{2005A&A...443...79B} Bastian, N., Gieles, M., Efremov, Y.~N., \& Lamers, H.~J.~G.~L.~M.\ 2005, A\&A, 443, 79 


\bibitem[Bastian et 
al.(2006)]{2006A&A...448..881B} Bastian, N., Saglia, R.~P., Goudfrooij, P., et al.\ 2006, A\&A, 448, 881 (B06)

\bibitem[Bastian et al.(2009)]{2009ApJ...701..607B} Bastian, N., Trancho, 
G., Konstantopoulos, I.~S., \& Miller, B.~W.\ 2009, ApJ, 701, 607 

\bibitem[Campbell et al.(2010)]{2010MNRAS.405..421C} Campbell, M.~A., 
Evans, C.~J., Mackey, A.~D., et al.\ 2010, MNRAS, 405, 421 

\bibitem[Efremov 
\& Elmegreen(1998)]{1998MNRAS.299..588E} Efremov, Y.~N., \& Elmegreen, B.~G.\ 1998, MNRAS, 299, 588 

\bibitem[Elson et al.(1987)]{1987ApJ...323...54E} Elson, R.~A.~W., Fall, 
S.~M., \& Freeman, K.~C.\ 1987, ApJ, 323, 54 (EFF)

\bibitem[Fellhauer 
\& Kroupa(2005)]{2005MNRAS.359..223F} Fellhauer, M., \& Kroupa, P.\ 2005, MNRAS, 359, 223 

\bibitem[Goudfrooij et al.(2001)]{2001MNRAS.328..237G} Goudfrooij, P., 
Alonso, M.~V., Maraston, C., \& Minniti, D.\ 2001, MNRAS, 328, 237 


\bibitem[Goudfrooij(2012)]{2012ApJ...750..140G} Goudfrooij, P.\ 2012, ApJ, 
750, 140 

\bibitem[Ha{\c s}egan et al.(2005)]{2005ApJ...627..203H} Ha{\c s}egan, M., 
Jord{\'a}n, A., C{\^o}t{\'e}, P., et al.\ 2005, ApJ, 627, 203 

\bibitem[King(1962)]{1962AJ.....67..471K} King, I.\ 1962, AJ, 67, 471 

\bibitem[Kissler-Patig et 
al.(2006)]{2006A&A...448.1031K} Kissler-Patig, M., Jord{\'a}n, A., \& Bastian, N.\ 2006, A\&A, 448, 1031 

\bibitem[Kotulla et al.(2009)]{2009MNRAS.396..462K} Kotulla, R., Fritze, 
U., Weilbacher, P., \& Anders, P.\ 2009, MNRAS, 396, 462 

\bibitem[Larsen(1999)]{1999A&AS..139..393L} Larsen, S.~S.\ 1999, A\&A, 139, 393 

\bibitem[Larsen et al.(2001)]{2001ApJ...556..801L} Larsen, S.~S., Brodie, 
J.~P., Elmegreen, B.~G., et al.\ 2001, ApJ, 556, 801 

\bibitem[Larsen et al.(2002)]{2002ApJ...567..896L} Larsen, S.~S., Efremov, 
Y.~N., Elmegreen, B.~G., et al.\ 2002, ApJ, 567, 896 

\bibitem[Larsen(2004)]{2004A&A...416..537L} Larsen, S.~S.\ 2004, A\&A, 416, 537 

\bibitem[Mackey 
\& Gilmore(2003)]{2003MNRAS.338...85M} Mackey, A.~D., \& Gilmore, G.~F.\ 2003, MNRAS, 338, 85 

\bibitem[Ma{\'{\i}}z-Apell{\'a}niz(2001)]{2001ApJ...563..151M} 
Ma{\'{\i}}z-Apell{\'a}niz, J.\ 2001, ApJ, 563, 151 

\bibitem[Maraston et al.(2004)]{2004A&A...416..467M} Maraston, C., Bastian, N., Saglia, R.~P., et al.\ 2004, A\&A, 416, 467 (M04)



\bibitem[Miller et al.(1997)]{1997AJ....114.2381M} Miller, B.~W., Whitmore, 
B.~C., Schweizer, F., \& Fall, S.~M.\ 1997, AJ, 114, 2381 (M97)

\bibitem[Moffat et al.(1994)]{1994ApJ...436..183M} Moffat, A.~F.~J., 
Drissen, L., \& Shara, M.~M.\ 1994, ApJ, 436, 183 

\bibitem[Portegies Zwart et 
al.(2010)]{2010ARA&A..48..431P} Portegies Zwart, S.~F., McMillan, S.~L.~W., \& Gieles, M.\ 2010, ARAA, 48, 431 

\bibitem[Rubin et al.(1970)]{1970ApJ...160..801R} Rubin, V.~C., Ford, W.~K., \& D'Odorico, S.\ 1970, ApJ, 160, 801

\bibitem[San Roman et al.(2012)]{2012MNRAS.426.2427S} San Roman, I., 
Sarajedini, A., Holtzman, J.~A., \& Garnett, D.~R.\ 2012, MNRAS, 426, 2427 

\bibitem[Schweizer(1982)]{1982ApJ...252..455S} Schweizer, F.\ 1982, ApJ, 
252, 455 

\bibitem[Schweizer \& Seitzer(1998)]{1998AJ....116.2206S} Schweizer, F., \& Seitzer, P.\ 1998, AJ, 116, 2206 (S98)

\bibitem[Schweizer(2004)]{2004ASPC..322..111S} Schweizer, F.\ 2004, The 
Formation and Evolution of Massive Young Star Clusters, 322, 111 

\bibitem[von Hoerner(1957)]{1957ApJ...125..451V} von Hoerner, S.\ 1957, ApJ, 125, 451 

\bibitem[Walborn \& Blades(1997)]{1997ApJS..112..457W} Walborn, N.~R., \& Blades, J.~C.\ 1997, ApJS, 112, 457 

\bibitem[Whitmore et al.(1993)]{1993AJ....106.1354W} Whitmore, B.~C., Schweizer, F., Leitherer, C., et al.\ 1993, AJ, 106, 1354

\bibitem[Whitmore et al.(1999)]{1999AJ....118.1551W} Whitmore, B.~C., Zhang, Q., Leitherer, C., et al.\ 1999, AJ, 118, 1551 

\bibitem[Whitmore et al.(2010)]{2010AJ....140...75W} Whitmore, B.~C., 
Chandar, R., Schweizer, F., et al.\ 2010, AJ, 140, 75 

\bibitem[Zhang et al.(2001)]{2001ApJ...561..727Z} Zhang, Q., Fall, S.~M., 
\& Whitmore, B.~C.\ 2001, ApJ, 561, 727 



\end{thebibliography}
\end{document}